\documentclass[aps,prx,twocolumn,footinbib,superscriptaddress,floatfix]{revtex4}

\usepackage{color}
\usepackage{epsfig}
\usepackage{latexsym}
\usepackage{amssymb}
\usepackage{amsmath}
\usepackage{algorithm}
\usepackage{algorithmic}
\usepackage{latexsym}
\usepackage{fancyhdr}
\usepackage{verbatim}
\usepackage{tabularx}
\usepackage{graphicx}
\usepackage{wasysym}
\usepackage{setspace}

\usepackage{psfrag,graphicx}
\usepackage{dcolumn}
\usepackage{bm}
\usepackage{amsfonts,amssymb,amsmath}
\usepackage{epstopdf}
\usepackage{amsthm}

\usepackage{times}

\renewcommand{\vec}[1]{\mathbf{#1}}

\newcommand{\plaq}{\square}

\newcommand{\ket}[1]{\left | \, #1 \right\rangle}
\newcommand{\bra}[1]{\left \langle #1 \, \right |}

\begin{document}

\title{Error Thresholds for Abelian Quantum Double Models:\\ 
Increasing the bit-flip Stability of Topological Quantum Memory}

\author{Ruben S.~Andrist}
\affiliation{Santa Fe Institute, 1399 Hyde Park Road, Santa Fe, NM
87501}

\author{James R.~Wootton}
\affiliation{Department of Physics, University of Basel,
Klingelbergstrasse 82, CH-4056 Basel, Switzerland}

\author{Helmut G.~Katzgraber}
\affiliation{Department of Physics and Astronomy, Texas A\&M
University, College Station, Texas 77843-4242, USA}
\affiliation{Materials Science and Engineering Program, Texas A\&M
University, College Station, Texas 77843, USA}
\affiliation{Santa Fe Institute, 1399 Hyde Park Road, Santa Fe, NM
87501}

\date{\today}

\begin{abstract}

    Current approaches for building quantum computing devices focus on
    two-level quantum systems which nicely mimic the concept of a
    classical bit, albeit enhanced with additional quantum properties.
    However, rather than artificially limiting the number of states to
    two, the use of $d$-level quantum systems (qudits) could provide
    advantages for quantum information processing. Among other merits,
    it has recently been shown that multi-level quantum systems can
    offer increased stability to external disturbances --- a key problem
    in current technologies. In this study we demonstrate that
    topological quantum memories built from qudits, also known as
    abelian quantum double models, exhibit a substantially increased
    resilience to noise. That is, even when taking into account the
    multitude of errors possible for multi-level quantum systems,
    topological quantum error correction codes employing qudits can
    sustain a larger error rate than their two-level counterparts. In
    particular, we find strong numerical evidence that the thresholds of
    these error-correction codes are given by the hashing bound.
    Considering the significantly increased error thresholds attained,
    this might well outweigh the added complexity of engineering and
    controlling higher dimensional quantum systems.

\end{abstract}

\maketitle

\section{Introduction}

The prospect of tremendous speedup over classical computation has
channeled considerable research effort into the pursuit of building a
universal quantum computer \cite{feynman:85,shor:97}. However, one of
the most significant challenges has been error correction: Despite
numerous attempts to physically realize and control quantum bits, all
share a striking sensitivity to decoherence -- external disturbance
attributed to the unavoidable interaction with the environment. And
while correcting these errors is in principle possible
\cite{shor:95,steane:96}, achieving error resilience in an efficient
and, in particular, scalable way is challenging and consequently a topic
of considerable research interest. 

Traditionally, quantum information processing is conceptualized as a
quantum adaptation of the binary storage system found in current
computers.  However, there are several indicators that ``thinking
outside the binary box'' might be a key ingredient to further progress
towards reliable quantum computation.  Recent results suggest hat
$d$-level quantum systems, so called ``qudits,'' are potentially more
powerful than their binary counterparts in terms of information
processing \cite{bechmann:00,nielsen:02,lanyon:09}.
Furthermore, while single-qubit gates can typically be implemented with
relatively high fidelity, two-qubit gates are often more challenging
because they require concurrent control over several parts of the
system.  Higher-dimensional quantum systems can allow for
information-coding with increased density, thus potentially reducing the
number of error-prone inter-qudit interactions required to perform
specific computations \cite{bullock:05}.

Different working quantum computing devices are currently being
scrutinized by a myriad of research teams around the globe: These range
from commercial quantum annealing devices, such as D-Wave Inc.'s D-Wave
Two quantum annealer \cite{comment:d-wave} based on superconducting flux
qubits, to proof-of-principle trapped-ion qubits that recently
demonstrated successfully fault-tolerant quantum computation based on
topological protection \cite{nigg:14}. While the former is built for
scalability, it suffers from decoherence effects mostly due to $1/f$
noise. The latter, on the other hand, is robust against noise due to the
inherent topological protection, however shows little prospect of
scalability.  Several of the current approaches to physically represent
quantum information can be extended to more than two levels, including
optical system \cite{mair:01,piani:11}, superconductors and flux
qubits \cite{neeley:09,comment:d-wave}, as well as atomic spins
\cite{klimov:03,mischuck:12}.  Given the theoretical promise that
qudit-based systems have shown
\cite{bechmann:00,nielsen:02,lanyon:09}, it is thus of
paramount interest to quantify their resilience to noise.

In terms of achieving error resilience for a given set of building
blocks, topological error-correction codes are currently among the best
candidates for a scalable error-correction scheme
\cite{kitaev:03,bombin:06}. This approach of using topology to reliably
encode quantum information can also be extended to multi-level qudit
systems, resulting in the so-called abelian quantum double models
\cite{bullock:07}. We are interested in estimating the dependence of the
error threshold on the qudit dimensionality -- that is, the maximum
amount of errors a system of $d$-level qudits can sustain while still
being able to reliably store the encoded information.  In the spirit of
the approach used to estimate the error tolerance of fault-tolerant
systems pioneered in Refs.~\cite{dennis:02,katzgraber:10,andrist:10} for
qubit systems, we present a mapping of this statistical analysis to
classical Potts-spin models that can be seen as the natural extension of
Boolean variables to $d$ states. This allows us to numerically
calculate the bit-flip error threshold for qudit-based quantum memories
using Monte Carlo simulations, a key figure of merit.

The paper is structured as follows. In Sec.~\ref{sec:abel} we introduce
abelian quantum double models, followed by both upper and lower
theoretical bounds for the threshold. A mapping onto a statistical
mechanical Potts model is outlined in Sec.~\ref{sec:model}, followed by
results in Sec.~\ref{sec:results} and concluding remarks.

\section{Abelian quantum double models}
\label{sec:abel}

The abelian quantum double models are defined on a square lattice with a
$d$-level quantum spin on each edge $\ell$. Let us use $\ket{s_\ell} \in
\{ \ket{0}, \ldots, \ket{d-1} \}$ to denote the computational basis of
the spins. We will use these to define a qudit generalization of the
toric code based on the cyclic group ${\mathbb Z}_d$, which we call the
$D({\mathbb Z}_d)$ code \cite{kitaev:03}.

\begin{figure}
    \begin{center}
    \includegraphics[width=8.5cm]{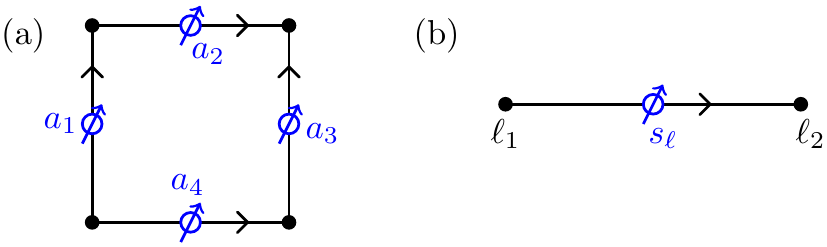}
    \caption{
        \label{fig:numbers}
	Numbering and orientation conventions used in the calculations.
	The quantum spins of the original model are shown as (blue)
	empty circles on the graph's edges, whereas the sites of the
	classical model are shown as (black) solid dots on the vertices.
	(a) Convention for determining anyon occupancy of a plaquette
	$\plaq$ from the quantum spins on the edges.  (b) Convention for
	assigning classical spins $S_{\ell_1}$ and $S_{\ell_2}$ to the
	vertices and determining the effective $\kappa_\ell$ for a given
	link $\ell$.
    }
	\vspace{-2mm}
    \end{center}
\end{figure}

Like the qubit-based toric code, operators are defined on the spins
around each plaquette and vertex of a square lattice in order to define
occupancies of anyonic quasiparticles. For the plaquettes, these
operators depend on relations between the states of the spins around the
plaquette when expressed in the computational basis. In the case that
all the spins $\plaq_j$ around a plaquette $\plaq$ are in definite
computational basis states $\ket{s_{\plaq_j}}$, the plaquette is said to
hold an anyon of type $n_{\plaq}$, where
\begin{equation}
    n_{\plaq} = (s_{\plaq_1} + s_{\plaq_2} - s_{\plaq_3} - s_{\plaq_4}) 
    \mod d\,.
\label{eq:toothbrush}
\end{equation}
The numbering of the spins in Eq.~\eqref{eq:toothbrush} around the
plaquette is depicted in Fig.~\ref{fig:numbers}(a). The $n=0$ anyon is
identified with the vacuum, while the $d-1$ other anyon types are
non-trivial. Note that many different states for the spins around a
plaquette lead to the same anyon type contained within. The typical
states of the system with a given anyon configuration will correspond to a
superposition of many of these.

The plaquette anyons can be created, moved, and annihilated using the
operations
\begin{equation}
    (\sigma^x_\ell)^{\epsilon} = \sum_{s_\ell} \ket{s_\ell+\epsilon\mod
    d}\bra{s_\ell}.
\end{equation}
There are $d$ distinct operations given by  $\epsilon \in \{ 0, \ldots,
d-1 \}$.  All are non-trivial except $\epsilon=0$, which yields the
identity.

For the vertices of the code anyon occupancies are defined in a similar
manner as for the plaquettes. The basis used is not the computational basis,
but that defined by
\begin{equation}
	\ket{\tilde j} = \frac{1}{\sqrt{d}} \sum_{k=0}^{d-1} e^{i 2 \pi jk/d}   \ket{j}.
\end{equation}
There are also corresponding operations $(\sigma^z_\ell)^{\epsilon}$
that manipulate these anyons. These vertex anyons are completely dual to
the plaquette ones, and so can be treated independently and analogously.
We therefore restrict our attention to plaquette anyons without loss of
generality.

Quantum information is stored in the vacuum states of the anyons, i.e.,
states where no anyons are present on any plaquette or vertex. There are
$d$ such states, allowing a logical qudit to be stored. The action of
any errors on the code can be expressed in terms of the
$(\sigma^x_\ell)^{\epsilon}$ and $(\sigma^z_\ell)^{\epsilon}$
operations, and so correspond to the creation and manipulation of
anyons. The resulting anyon configuration, which serves as the syndrome
of the code, then allows information about the errors to be determined.
The goal of error correction is to use this information to remove the
effects of the errors.

Error correction will not always succeed. The probability with which it
fails depends on the linear system size of the code, $L$, and the
strength of the noise, $p$. If $p$ is below a threshold value, which we
denote $p_d$ for a $D({\mathbb Z}_d)$ code, the probability of failure
decays exponentially with $L$. Success then become certain as $L
\rightarrow \infty$. For $p > p_d$, however, the probability of failure
is always $O(1)$.

To determine the value of the error threshold $p_d$ for a $D({\mathbb
Z}_d)$ code, we must consider a specific error model. Here we consider a
generalization of the well-known independent bit and phase flip error
model for the ${\mathbb Z}_2$ case \cite{dennis:02}. This acts
independently on each spin of the code, applying a randomly chosen
non-trivial operation $(\sigma^x_\ell)^{\epsilon}$ with probability $p$.
The probability of each non-trivial operation is therefore $p/(d-1)$.
See Refs.~\cite{duclos:13,anwar:13} for the best efforts to perform
error correction for this problem to date.

In Sec.~\ref{sec:results}, the thresholds are determined numerically for several
different values of $d$. However it is interesting to consider how $p_d$ may
behave for $d\rightarrow \infty$.

For the ${\mathbb Z}_2$ code it has been observed in multiple cases that
the threshold follows the Hashing bound
\cite{roethlisberger:11,bombin:12}.  It has been speculated that this
may also hold in the qudit case \cite{duclos:13,anwar:13}. For the error
model we consider, the Hashing bound probability $p_{\rm hb}$ is the
solution to the equation
\begin{equation}
\label{eq:hashingbound}
p_{\rm hb}\log(d-1) - p_{\rm hb}\log p_{\rm hb} 
- (1-p_{\rm hb})\log(1-p_{\rm hb}) 
= \frac{\log d}{2}\,.
\end{equation}
Note that although this is referred to as a ``bound,'' it serves as an
{\em upper} bound only for non-degenerate codes. Because the codes we
consider are degenerate, they are not necessarily limited by these
values.

\vspace{-1mm}
\subsection{Upper bound for thresholds}

Even though the Hashing bound is not a strict upper bound for the codes
we consider, such a bound can be derived. Suppose all spins of a
$D({\mathbb Z}_d)$ code are prepared in state $\ket{0}$. If the anyon
configuration is then measured, it will be found that there is only
vacuum on all plaquettes but a random pattern of anyons on the vertices.
When the latter are removed, the resulting state of the code is that for
which the logical qubit state is $\ket{0}$.

Now suppose that a unitary $U(\theta)$ is applied to all spins prior
to the measurement, such that
\begin{equation}
U(\theta) \ket{0} 
= \cos \theta \ket{0} 
+ \sin \theta \frac{1}{\sqrt{d-1}} \sum_{j=1}^{d-1} \ket{j}.
\end{equation}
This has the same effects on the measurement results as applying errors
of the form $(\sigma^x)^g$ for $1 \leq g \leq d-1$ with equal
probability $p/(d-1)$, where $p=\sin^2 \theta$. The threshold $p_d$ in
the latter case therefore corresponds to a threshold angle $\theta_d =
\arcsin \sqrt p_d$ in the former. For $\theta < \theta_d$ error
correction will yield the code in logical state $\ket{0}$ with certainty
as $L \rightarrow \infty$.

A similar argument can be applied using the complementary basis in which
the vertex anyon states are defined. If each qudit is prepared in state
$\ket{\tilde 0}$, syndrome measurement and error correction result
in the logical qubit state $\ket{\tilde 0}$. Applying the unitary operator
\begin{equation}
V(\phi) \ket{\tilde 0} = 
	\cos \phi \ket{\tilde 0} 
	+ \sin \phi \frac{1}{\sqrt{d-1}} \sum_{j=1}^{d-1} \ket{\tilde j}
\end{equation}
to each spin before measurement has the same effect as errors of the
form $(\sigma^z)^g$ for $1 \leq g \leq d-1$ with equal probability
$p/(d-1)$ where $p=\sin^2 \phi$. There will then be a critical angle
$\phi_d = \arcsin \sqrt p_d$ below which error correction yields the
code in logical state $\ket{\tilde 0}$ with certainty as $L \rightarrow
\infty$.

Note that, for particular values of $\theta$ and $\phi$,
\begin{equation}
U(\theta) \ket{0} = 
		V(\phi) \ket{\tilde 0} 
		  = 
		\sqrt{ \frac{d}{ 2 (d + \sqrt{d})}}  
		\left( \ket{0} + \ket{\tilde 0} \right) \, .
\end{equation}
This corresponds to both of the above cases with the errors occurring
with probability
\begin{equation}
	p = \frac{d-1}{2(d+\sqrt{d})}.
\end{equation}
As such, error correction cannot successfully correct both types of
error, because it cannot yield a code whose logical state is
simultaneously $\ket{0}$ and $\ket{\tilde 0}$. This error rate must
therefore be above the threshold, giving us an upper bound
\begin{equation}
	p_d < \frac{d-1}{2(d+\sqrt{d})}
\end{equation}
for the threshold error rate $p_d$ of the $D({\mathbb Z}_d)$ code.

\subsection{Lower bound for thresholds}

For some values of $d$ it is also possible to construct a lower bound:
Given an integer $d=nm$ where $n$ and $m$ are co-prime, the fundamental
theorem of abelian groups states that the group ${\mathbb Z}_d$ is
isomorphic to ${\mathbb Z}_n \times {\mathbb Z}_m$.  As such, the
problem of decoding an error correcting code based on the $D({\mathbb
Z}_d)$ quantum double model is the same as that for two separate codes,
one based on $D({\mathbb Z}_n)$ and the other on $D({\mathbb Z}_m)$. We
may therefore consider the independent decoding of these two component
codes, rather than decoding the full $D({\mathbb Z}_d)$ code.

By definition, the optimal decoding of the component codes cannot do
better than optimal decoding for the full $D({\mathbb Z}_d)$ code.  In
general it will do worse, because the error model for the full
$D({\mathbb Z}_d)$ code will translate to one on the component codes
with strong correlations between the two. Ignoring these correlations by
decoding the component codes independently leads to decreased
performance.

For the $D({\mathbb Z}_d)$ code there are $d-1$ non-trivial errors of the
form $(\sigma^x)^g$, corresponding to $0 < g \leq d-1$. The number of
these that act non-trivially on the ${\mathbb Z}_n$ code is $(n-1)m$:
the number of elements of ${\mathbb Z}_n \times {\mathbb Z}_m$ that do
not correspond to the identity for ${\mathbb Z}_n$. These all occur with
equal probability $p/(d-1)$. Using $p^{(n)}$ to denote the total
probability of error for the ${\mathbb Z}_n$ code, it follows
\begin{equation}
    \nonumber
    \frac{p^{(n)}}{(n-1)m} = \frac{p}{d-1}, \,\,\,\, 
        \therefore \,\,\,\, p = p^{(n)} \, \frac{d-1}{d-m}\,.
\end{equation}
When decoding the component codes independently, error correction fails
when either of the component codes experience an error rate above their
threshold. The error rate $p$ at which this occurs is then a lower bound
for $p_d$, i.e., 
\begin{equation} 
    \nonumber
	p_d \geq  \min \left(   
		p_n \, \frac{d-1}{d-m}  ,   p_m \, \frac{d-1}{d-n}   
	\right)
    \stackrel{n<m}{\geq}
    p_n\,\frac{d-1}{d-n}\,.
\end{equation}
We can also use this as a self-consistency check for any proposed set of
thresholds $p_d^{\rm prop}$, such as those provided by the hashing
bound, Eq.~\eqref{eq:hashingbound}. The easiest way to do this is to
use prime $n$, ensuring that $m=n+1$ is co-prime. Self-consistency then
requires
\begin{equation} 
    \nonumber 
	p_{n^2+n}^{\rm prop} \geq p_n^{\rm prop} \, \left( 
		1 + \frac{1}{n} - \frac{1}{n^2} 
	\right)\,,
\end{equation}
which is indeed satisfied by the hashing bound values.

\section{Mapping and Numerical Simulations}
\label{sec:model}

We first map the problem of computing the error threshold to a classical
statistical-mechanical model with disorder.  The threshold where errors
in the quantum setup cannot be corrected any longer will correspond to
the loss of a symmetry broken phase in the classical model.

\subsection{Mapping}

For the mapping we consider a square lattice and an initially anyon-free
state, $n_{\plaq} = 0\,\forall \plaq$. Errors
$(\sigma^x_\ell)^{\epsilon_\ell}$ are applied independently to each
quantum spin, where $\epsilon_\ell$ denotes the type of error applied to
the spin on edge $\ell$. We initially perform the mapping with a general
error model where each value occurs with respective probability
$p_\epsilon$ and the error is non-trivial with total probability $p =
\sum_{\epsilon\neq0}p_\epsilon$.  The probability for a given error
configuration $E = \{\epsilon_l\}$ is then
\begin{align} 
	\label{eq:general_prob}
	P(E) &= \prod_\ell \prod_\epsilon p_\epsilon^{\delta_{\epsilon_\ell,\epsilon}}
	 = \prod_\ell \exp [ 
		 -\beta \sum_\epsilon J_\epsilon \delta_{\epsilon_\ell,\epsilon}
	 ]\,,
\end{align}
where $\beta$, $J_\epsilon$ and $p_\epsilon$ satisfy
\begin{equation}
    \label{eq:betaJrelation}
    p_{\epsilon} = e^{-\beta J_{\epsilon}}\,.
\end{equation}
Assuming an initially anyon free state, the set of errors $E$ yields an
anyon configuration with the type of the anyon on plaquette $\plaq$
given by
\begin{equation}
    \nonumber
    n^E_{\plaq} = \epsilon_{\plaq_1} + \epsilon_{\plaq_2} 
        - \epsilon_{\plaq_3} - \epsilon_{\plaq_4}
	\mod d\,.
\end{equation}
Let us now consider another set of errors $E' = \{\epsilon'_\ell\}$.
This can be related to $E$ by a third set $C = \{\kappa_\ell\}$
according to
\begin{equation}
    \nonumber
	\label{eq:general_diff}
	\epsilon'_\ell = \epsilon_\ell + \kappa_\ell \mod d\,.
\end{equation}
Note that if $E$ and $E'$ lead to the same anyon configuration,
$n^E_{\plaq} = n^{E'}_{\plaq}$, then the anyon configuration of $C$ is 
be trivial. Explicitly, it satisfies
\begin{equation}
	\label{eq:no_anyons}
	n^C_{\plaq} = (\kappa_{\plaq_1} + \kappa_{\plaq_2} - \kappa_{\plaq_3} 
		- \kappa_{\plaq_4}) \mod d \, = 0 \,\,\, \forall \, \plaq\,.
\end{equation}
This follows from the fact that $\kappa_\ell = \epsilon'_\ell -
\epsilon_\ell$, and hence $n^C_{\plaq} = n^{E'}_{\plaq} - n^E_{\plaq} =
0$ (both evaluated $\mod d$). Combining Eqs.~\eqref{eq:general_prob}
and~\eqref{eq:general_diff}, we find that the ratio of the probabilities
for $E'$ and $E$ can be written as
\begin{equation} 
    \nonumber
	\frac{P(E')}{P(E)} = \prod_\ell \exp [ 
        -\beta \sum_\epsilon J_\epsilon(
            \delta_{\epsilon_\ell+\kappa_\ell,\epsilon} 
            - \delta_{\epsilon_\ell,\epsilon}
        ) 
	]\,,
\end{equation}
where the sum $\epsilon_\ell+\kappa_\ell$ is again evaluated $\mod d$.
This ratio can be interpreted as a Boltzmann weight for a classical
model with the Hamiltonian
\begin{equation}
    \label{eq:nonspin_hamiltonian}
    \mathcal{H} = \sum_\ell\sum_\epsilon J_\epsilon(
          \delta_{\epsilon_\ell+\kappa_\ell,\epsilon}
	- \delta_{\epsilon_\ell,\epsilon}
    )\,.
\end{equation}
We interpret the $\{\kappa_\ell\}$ as the state of the system and
$\{\epsilon_\ell\}$ as representing the details of the (quenched)
interactions. Because the partition function sums only over the former,
the second term in the brackets of Eq.~\eqref{eq:nonspin_hamiltonian} is
an overall constant and hence can be ignored.

To ensure that $E$ and $E'$ belong to the same error class, we require
that a system state $\{\kappa_\ell\}$ satisfies the constraint in
Eq.~\eqref{eq:no_anyons}. This is achieved by introducing an auxiliary
$d$-level classical spin $S_v \in \{ 0, \ldots, d-1 \}$ on each vertex
$v$ of the lattice. The $\kappa_\ell$ can then be defined by
$\kappa_\ell = S_{\ell_2} - S_{\ell_1} $ with the numbering as indicated
in Fig.~\ref{fig:numbers}(b). All the $S_v$ around a plaquette $\plaq$
will then appear exactly twice in $n^C_{\plaq}$ (and with opposite
sign), thus leaving the required value of zero in all cases.

In summary, we can sample from the set of errors $E'$ with the same
anyon configuration as a set $E$ (i.e., from the {\em error class} of
$E$) by sampling from a classical two-dimensional statistical-mechanical
spin model defined on a square lattice with $N = L\times L$ sites and
$d$-level spins on the vertices. The classical spin model of interest
has the Hamiltonian
\begin{equation}
	\label{eq:general_hamiltonian}
    \mathcal{H}(E) = \sum_\ell\sum_\epsilon J_\epsilon \delta_{\epsilon_\ell+
        S_{\ell_1} - S_{\ell_2} ,\epsilon}\,,
\end{equation}
where the errors of $E=\{\epsilon_\ell\}$ are generated according to the
error model. The inverse temperature $\beta$ and the interaction constants $J_\epsilon$ need to be scaled
accordingly to satisfy Eq.~\eqref{eq:betaJrelation}.

\begin{figure}[t]
    \begin{center}
	\vspace{1mm}
    \includegraphics[width=8.6cm]{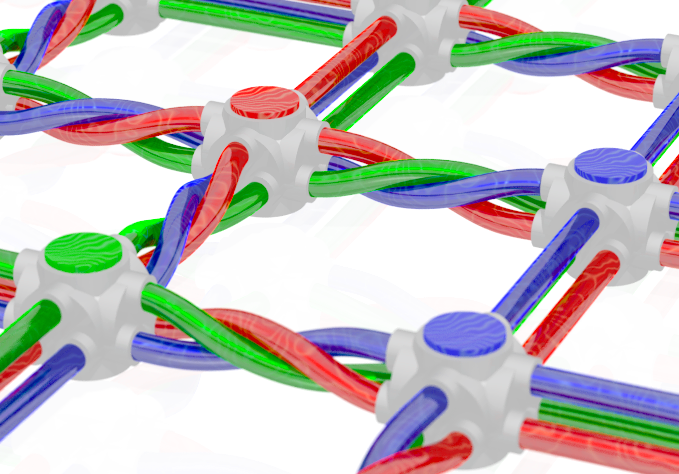}
    \vspace{-4mm}
    \caption{(Color online)
	Rendering of the classical three-colored Potts model found in
	the mapping of an abelian quantum double model with qutrits
	(i.e., qubits with $d = 3$) on each edge.  Quantum errors give
	rise to disorder in the interactions between neighboring sites,
	indicated in the form of twisted links. An ordered state in this
	model (i.e., one predominant color in the system) in spite of
	these faulty links indicates resilience of the original setup to
	quantum errors.
    }
    \label{fig:rendered} 
	\vspace{-6mm}
    \end{center}
\end{figure}

For the purpose of our numerical simulations, we consider the special
case where all non-trivial errors have equal probability: $p_\epsilon =
p/(d-1)\,\,\,\forall \epsilon\neq0$. In this case the simplest way to satisfy
Eq.~\eqref{eq:betaJrelation} is to sample at inverse temperature
\begin{equation}
    \label{eq:nishimori}
    \beta = -\ln\frac{p/(d-1)}{1-p}\,.
\end{equation}
This is the generalization of the Nishimori
condition~\cite{nishimori:81}, which, for the symmetrical case, ensures
that the Boltzmann weight of each state corresponds to the a priori
probability of the quantum error configuration it represents. The spin
Hamiltonian, up to an irrelevant overall energy shift, then takes the
form
\begin{equation}
    \label{eq:hamiltonian}
    \mathcal{H}(E) = -\sum_\ell \delta_{\epsilon_\ell+
        S_{\ell_q} - S_{\ell_2},0}\,,
\end{equation}
which can be interpreted as a disordered $d$-states Potts model (see
Fig.~\ref{fig:rendered} for a graphical illustration). In the mapping
each quantum error is translated to a non-trivial interaction constant
$\epsilon_\ell$ and the difference, $C$, is associated with the surface
of a flipped domain of Potts spins. Therefore, if in our sampling we
find the system in an overall ordered state, there is little uncertainty
in the type of error that occurred. If by contrast the system is
disordered, then the domain walls have percolated the system and we are
unable to correctly identify the class of the actual error.

\subsection{Numerical details: Algorithm \& Observables}

\paragraph*{Algorithm} --- To calculate the error threshold (i.e., where
ordering disappears), we numerically investigate
Eq.~\eqref{eq:hamiltonian} via large-scale (classical) Monte Carlo
simulations using the parallel tempering technique \cite{hukushima:96}.
Several replicas of the system are simulated concurrently for the same
disorder realization, but at different temperatures. In addition to
local Metropolis updates \cite{newman:99}, one performs global moves in
which the temperatures of two neighboring copies are exchanged. It is
important to select the position of the individual temperatures
carefully such that the acceptance probabilities for the global moves
are large enough \cite{katzgraber:06a} and each copy performs a random
walk in temperature space. This, in turn, allows each copy to
efficiently sample the configuration space, therefore speeding up the
simulation several orders of magnitude.

\paragraph*{Observables} --- Detecting the transition temperature
$T_c(p)$ to an ordered (ferromagnetic) phase for different fixed amounts
of disorder $p$ allows us to pinpoint the phase boundary in the
$p\,$--$T$ phase diagram. The error threshold $p_c$ is then given by the
intersection of the phase boundary with the Nishimori line,
Eq.~\eqref{eq:nishimori}.  To detect ordering, we use the simplex
representation where the $d$ states of the Potts spins are mapped to the
corners of a hyper-tetrahedron in $(d-1)$ space dimensions. This means
they are represented as a $(d-1)$-component unit vector $\vec{S}_i$
taking one of $d$ possible values satisfying the condition
$\vec{S}^\mu\cdot \vec{S}^{\nu} = 1 - \delta_{\mu,\nu}[d/(d-1)]$ with
$\{\mu,\nu\} \in \{1,2, \ldots, d\}$.  
Within this mapping, the magnetic
susceptibility of the disordered Potts model in
Eq.~\eqref{eq:hamiltonian} can be computed via
\begin{equation}
    \chi({\bf k}) = \sum_{\mu} \left[\left\langle\left| 
        \textstyle\sum_i \vec{S}_i^{\mu} e^{i {\bf k} \cdot 
            {\bf R}_i}\displaystyle
    \right|^2\right\rangle\right]_{\rm av}\,,
    \label{eq:chi}
\end{equation}
where the sum is over all Potts spins $\vec{S}_i$,
$\langle\,\cdots\rangle$ denotes a thermal average and $[\,\cdots]_{\rm
av}$ is an average over disorder realizations.  The presence of a
transition is probed by studying the two-point finite-size correlation
length \cite{palassini:99b},
\begin{equation}
	\label{eq:correlation_function}
    \xi_L = \frac{1}{2\sin(k_\mathrm{min}/2)} 
        \sqrt{\frac{\chi({\bf 0})}{\chi({\bf k}_\mathrm{min})} - 1}\,,
\end{equation}
where $\mathbf{k}_{\rm min} = (2\pi/L,0)$ is the smallest nonzero wave
vector for the given lattice. Near the transition, $\xi_L$ is expected
to scale as
\begin{equation}
	\xi_L/L \sim \tilde X[L^{1/\nu}(T-T_c)]\,,
	\label{eq:scaling}
\end{equation}
where $\tilde X$ is a dimensionless scaling function. Because the
argument of Eq.~\eqref{eq:scaling} becomes zero at the transition
temperature (and hence independent of~$L$), we expect lines of different
system sizes to cross at this point. If however the lines do not meet,
we know that no transition occurs in the studied temperature range.
This approach has been successfully used before for qubit systems, see
for example, Ref.~\cite{katzgraber:09c}.

\paragraph*{Finite-size scaling} --- In practice, there are corrections
to scaling to Eq.~(\ref{eq:scaling}) and the data for different system
sizes do not cross exactly at one temperature $T_c$ as suggested by the
finite-size scaling form, Eq.~\eqref{eq:scaling}.  That is, the actual
crossing $T_c^*$ between a pair of system sizes $L_1$ and $L_2$ shifts
with increasing system size $L$ and tends to a constant for $L_1$ and
$L_2 \to\infty$. To estimate the proper thermodynamic value we study
$T_c^*(L_1,L_2)$ as a function of the average inverse system size,
$2/(L_1\!+\!L_2)$, and fit a linear function to the data. The intercept
with the vertical axis is our estimate for the transition temperature in
the thermodynamic limit. The error bars are determined via a bootstrap
analysis using $500$ resamplings.

\paragraph*{Thermalization} --- In all simulations, equilibration is
tested using a logarithmic binning of the data. Once the data for all
observables measured agree within error bars for three logarithmic bins
we deem the Monte Carlo simulation for that system size to be in thermal
equilibrium. The detailed simulation parameters are listed in table
\ref{tab:simparams}.

\begin{table}[!tb]
    \vspace{-2.2mm}
    \caption{
	Simulation parameters: $d$ is the qudit dimensionality, $p$ is
	the qudit error rate, $L$ is the linear system size, $N_{\rm
	sa}$ is the number of disorder samples, $t_{\rm eq} = 2^{b}$ is
	the number of equilibration sweeps (system size times number of
	single-spin Monte Carlo updates), $T_{\rm min}$ [$T_{\rm max}$]
	is the lowest [highest] temperature, and $N_{\rm T}$ the number
	of temperatures used.
    }
    \label{tab:simparams}
    \vspace*{1mm}
    \centering
    {\footnotesize
    \begin{tabular*}{8 cm}{@{\extracolsep{\fill}} c c c r r r r r}
    \hline
    \hline
    $d$ & $p$ & $L$ & $N_{\rm sa}$ & $b$ & 
        $T_{\rm min}$ & $T_{\rm max}$ &$N_{\rm T}$ \\
    \hline 
    $3,4$ & $0.00-0.13$ & $12,16$ & $10\,000$  & $17$ & $0.60$ & $1.40$ & $24$\\
    $3,4$ & $0.00-0.13$ & $24,32$ & $5\,000$   & $20$ & $0.60$ & $1.40$ & $28$\\
    $3,4$ & $0.00-0.13$ & $48,64$ & $500$      & $21$ & $0.60$ & $1.40$ & $36$\\
    $3,4$ & $0.15-0.19$ & $12,16$ & $10\,000$  & $19$ & $0.35$ & $1.30$ & $24$\\
    $3,4$ & $0.15-0.19$ & $24,32$ & $5\,000$   & $23$ & $0.35$ & $1.30$ & $42$\\
    $3,4$ & $0.15-0.19$ & $48,64$ & $500$      & $24$ & $0.35$ & $1.30$ & $64$\\
    $6,10$ & $0.00-0.20$ & $12,16$ & $10\,000$ & $17$ & $0.45$ & $1.40$ & $24$\\
    $6,10$ & $0.00-0.20$ & $24,32$ & $1\,000$  & $21$ & $0.45$ & $1.40$ & $42$\\
    $6,10$ & $0.21-0.26$ & $12,16$ & $10\,000$ & $19$ & $0.25$ & $1.30$ & $32$\\
    $6,10$ & $0.21-0.26$ & $24,32$ & $1\,000$  & $24$ & $0.25$ & $1.30$ & $56$\\
    \hline
    \hline
    \end{tabular*}
    }
\end{table}

\section{Results}
\label{sec:results}

\begin{figure}
	\vspace{-3mm}
    \hbox{\hspace{-7mm}\includegraphics[width=10cm]{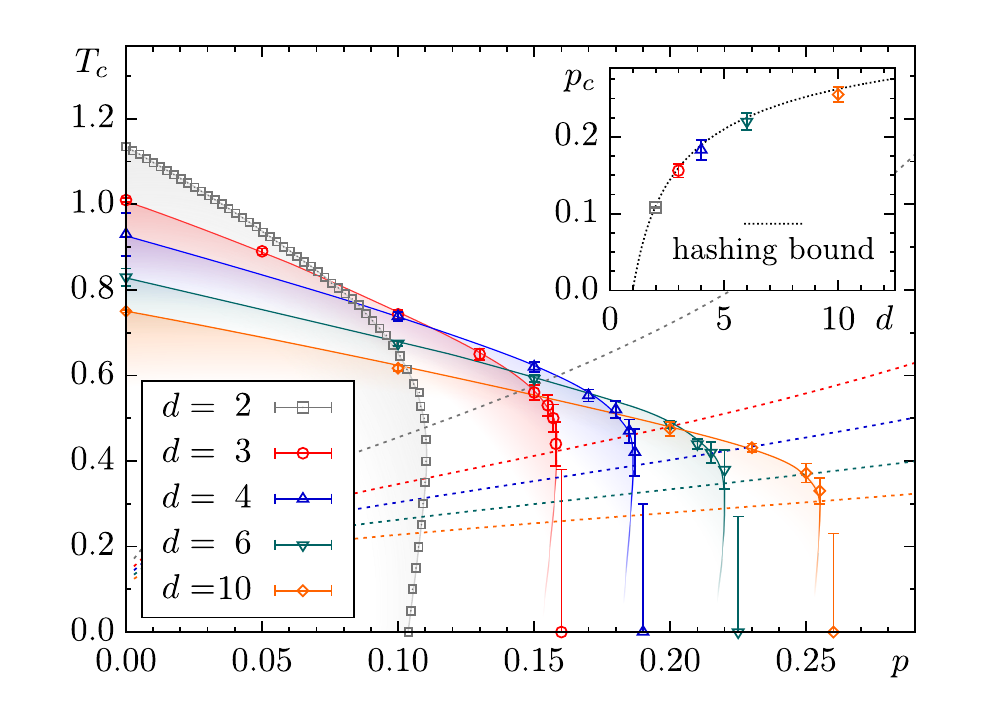}}
    \vspace{-3mm}
	\caption{(Color online)
	Phase boundaries for different $d$ estimated from Monte Carlo
	simulations of the respective disordered Potts models
	representing quantum error correction for $d$-level quantum
	systems. The solid lines and shaded areas are guides to the eye.
	The error
	threshold $p_c(d)$ corresponds to the point where the
	corresponding Nishimori lines (dashed) intersects the phase
	boundaries for different $d$. This is the maximum disorder rate
	for which error correction is feasible. Inset: Comparison of
	the calculated error thresholds for different values of $d$ to
	the upper bound given by the hashing bound (dotted black line).
	There is excellent agreement.
	}
	\label{fig:phasediagram}
\end{figure}

In the mapping, the special case of $d=2$ corresponds to the glassy
Ising spin systems studied in previous studies
\cite{dennis:02,thomas:11d}, albeit with a temperature that differs by a
factor of two. In Fig.~\ref{fig:phasediagram}, we show the detailed
results by Thomas {\em et~al.}~\cite{thomas:11d} rescaled appropriately
for comparison. Note that this setup is stable up to close to the
hashing bound value of $p_{\rm hb} \approx 0.11$, i.e., even if
approximately $10.9\%$ of the qubits are faulty, errors can still be
corrected.

For qutrits ($d=3$) without any disorder, the transition occurs at
$T_{c,d=3} \approx 1.01(4)$. As the amount of disorder is gradually
increased, the transition temperature $T_{c}(d,p)$ is lowered until it
meets the Nishimori line. Because finite-size effect are most pronounced
close to the critical error threshold, the error bars calculated by
bootstrapping the finite-size scaling analysis are larger. The largest
value for which we find a crossing in the two-point finite-size
correlation function is $p=0.158$. This means that even when $15.8\%$ of
the qutrits in a topological memory are faulty, error correction is
still possible and the encoded information is retained. Note that the
limited temperature ranges studied here do not allow us to detect a
potential reentrance effect as observed in Ref.~\cite{thomas:11d}. Our
conservative numerical estimate for the error threshold for qutrits,
$p_{c,d=3} = 0.158(2)$, agrees within error bars with the theoretical
limit given by the hashing bound, Eq.~\eqref{eq:hashingbound}.

For higher dimensional qudits ($d = 4$, $6$ and $10$), an analogous
analysis indicates that the error threshold for these codes actually
tracks the hashing bound to within the precision our numerical results.
This is shown in the inset of Fig.~\ref{fig:phasediagram}, which relates
the numerical estimates for the error thresholds (data points) to the
respective hashing bound value (line).  Note that the precision is
limited by the numerical sampling effort: The required equilibration
time for disordered $d$-level Potts systems increases dramatically for
larger values of $d$, because the configuration space grows
exponentially and the transition temperature is lowered at the same
time. However, we do emphasize that the topological stability of
qudit-based codes increases monotonically with $d$, i.e., implementing
higher-level qudits leads to increasingly stable topologically protected
memory.

\section{Conclusion}

We have demonstrated that error correction in topological memory built
from qudits is closely related to the ferromagnetic ordering in
classical disordered $d$-state Potts systems. By numerically estimating
the phase boundary for the related classical statistical-mechanical
model, we show that error correction remains feasible up to the hashing
bound (within error bars) for values of $d$ between $2$ and $10$. Our
results are summarized in Fig.~\ref{fig:phasediagram}, which shows the
estimated phase diagram and Nishimori line for all values of $d$
studied. The inset relates the numerical estimates for the error
thresholds to the hashing bound, Eq.~\eqref{eq:hashingbound}. In
particular our results demonstrate that, by moving from qubit to qutrit
building blocks, the resilience of a topological quantum memory can
potentially be increased already by at least 42\%. This is especially
encouraging because currently available technology, e.g., based on flux
qubits \cite{comment:d-wave} already allows for the encoding of
$d$-level quantum bits. In the particular case of the technology used in
the D-Wave Two quantum annealer, theoretically, qudits with $d$ values
up to $5$ can be implemented.

\begin{acknowledgments} 

The authors would like to thank the Santa Fe Institute for HPC resources
on the Scoville cluster, and in particular N.~Metheny for its 
administration and continued technical support.  H.G.K.~acknowledges
support from the National Science Foundation (Grant No.~DMR-1151387) and
would like to thank the Santa Fe Institute for their hospitality.
J.R.W.~acknowledges support from the Swiss National Science Foundation, 
as well as and NCCR QSIT.

\end{acknowledgments}


\begin{thebibliography}{32}
\expandafter\ifx\csname natexlab\endcsname\relax\def\natexlab#1{#1}\fi
\expandafter\ifx\csname bibnamefont\endcsname\relax
  \def\bibnamefont#1{#1}\fi
\expandafter\ifx\csname bibfnamefont\endcsname\relax
  \def\bibfnamefont#1{#1}\fi
\expandafter\ifx\csname citenamefont\endcsname\relax
  \def\citenamefont#1{#1}\fi
\expandafter\ifx\csname url\endcsname\relax
  \def\url#1{\texttt{#1}}\fi
\expandafter\ifx\csname urlprefix\endcsname\relax\def\urlprefix{URL }\fi
\providecommand{\bibinfo}[2]{#2}
\providecommand{\eprint}[2][]{\url{#2}}

\bibitem[{\citenamefont{Feynman}(1985)}]{feynman:85}
\bibinfo{author}{\bibfnamefont{R.}~\bibnamefont{Feynman}},
  \emph{\bibinfo{title}{{{Quantum mechanical computers}}}},
  \bibinfo{journal}{Optics news} \textbf{\bibinfo{volume}{11}},
  \bibinfo{pages}{1} (\bibinfo{year}{1985}).

\bibitem[{\citenamefont{Shor}(1997)}]{shor:97}
\bibinfo{author}{\bibfnamefont{P.~W.} \bibnamefont{Shor}},
  \emph{\bibinfo{title}{{{Polynomial-time algorithms for prime factorization
  and discrete logarithms on a quantum computer}}}}, \bibinfo{journal}{SIAM J.
  Comp.} \textbf{\bibinfo{volume}{26}}, \bibinfo{pages}{1484}
  (\bibinfo{year}{1997}).

\bibitem[{\citenamefont{Shor}(1995)}]{shor:95}
\bibinfo{author}{\bibfnamefont{P.~W.} \bibnamefont{Shor}},
  \emph{\bibinfo{title}{{{Scheme for reducing decoherence in quantum computer
  memory}}}}, \bibinfo{journal}{Phys. Rev. A} \textbf{\bibinfo{volume}{52}},
  \bibinfo{pages}{R2493} (\bibinfo{year}{1995}).

\bibitem[{\citenamefont{Steane}(1996)}]{steane:96}
\bibinfo{author}{\bibfnamefont{A.~M.} \bibnamefont{Steane}},
  \emph{\bibinfo{title}{{{Error Correcting Codes in Quantum Theory}}}},
  \bibinfo{journal}{Phys. Rev. Lett.} \textbf{\bibinfo{volume}{77}},
  \bibinfo{pages}{793} (\bibinfo{year}{1996}).

\bibitem[{\citenamefont{Bechmann-Pasquinucci and Peres}(2000)}]{bechmann:00}
\bibinfo{author}{\bibfnamefont{H.}~\bibnamefont{Bechmann-Pasquinucci}}
  \bibnamefont{and} \bibinfo{author}{\bibfnamefont{A.}~\bibnamefont{Peres}},
  \emph{\bibinfo{title}{{{Quantum cryptography with 3-state systems}}}},
  \bibinfo{journal}{Phys. Rev. Lett.} \textbf{\bibinfo{volume}{85}},
  \bibinfo{pages}{3313} (\bibinfo{year}{2000}).

\bibitem[{\citenamefont{Nielsen et~al.}(2002)\citenamefont{Nielsen, Bremner,
  Dodd, Childs, and Dawson}}]{nielsen:02}
\bibinfo{author}{\bibfnamefont{M.~A.} \bibnamefont{Nielsen}},
  \bibinfo{author}{\bibfnamefont{M.~J.} \bibnamefont{Bremner}},
  \bibinfo{author}{\bibfnamefont{J.~L.} \bibnamefont{Dodd}},
  \bibinfo{author}{\bibfnamefont{A.~M.} \bibnamefont{Childs}},
  \bibnamefont{and} \bibinfo{author}{\bibfnamefont{C.~M.}
  \bibnamefont{Dawson}}, \emph{\bibinfo{title}{{Universal simulation of
  Hamiltonian dynamics for quantum systems with finite-dimensional state
  spaces}}}, \bibinfo{journal}{Phys. Rev. A} \textbf{\bibinfo{volume}{66}},
  \bibinfo{pages}{022317} (\bibinfo{year}{2002}).

\bibitem[{\citenamefont{Lanyon et~al.}(2009)\citenamefont{Lanyon, Barbieri,
  Almeida, Jennewein, Ralph, Resch, Pryde, O'Brien, Gilchrist, and
  White}}]{lanyon:09}
\bibinfo{author}{\bibfnamefont{B.~P.} \bibnamefont{Lanyon}},
  \bibinfo{author}{\bibfnamefont{M.}~\bibnamefont{Barbieri}},
  \bibinfo{author}{\bibfnamefont{M.~P.} \bibnamefont{Almeida}},
  \bibinfo{author}{\bibfnamefont{T.}~\bibnamefont{Jennewein}},
  \bibinfo{author}{\bibfnamefont{T.~C.} \bibnamefont{Ralph}},
  \bibinfo{author}{\bibfnamefont{K.~J.} \bibnamefont{Resch}},
  \bibinfo{author}{\bibfnamefont{G.~J.} \bibnamefont{Pryde}},
  \bibinfo{author}{\bibfnamefont{J.~L.} \bibnamefont{O'Brien}},
  \bibinfo{author}{\bibfnamefont{A.}~\bibnamefont{Gilchrist}},
  \bibnamefont{and} \bibinfo{author}{\bibfnamefont{A.~G.} \bibnamefont{White}},
  \emph{\bibinfo{title}{{{Simplifying quantum logic using higher-dimensional
  Hilbert spaces}}}}, \bibinfo{journal}{Nat. Phys.}
  \textbf{\bibinfo{volume}{5}}, \bibinfo{pages}{134} (\bibinfo{year}{2009}).

\bibitem[{\citenamefont{Bullock et~al.}(2005)\citenamefont{Bullock, O'Leary,
  and Brennen}}]{bullock:05}
\bibinfo{author}{\bibfnamefont{S.~S.} \bibnamefont{Bullock}},
  \bibinfo{author}{\bibfnamefont{D.~P.} \bibnamefont{O'Leary}},
  \bibnamefont{and} \bibinfo{author}{\bibfnamefont{G.~K.}
  \bibnamefont{Brennen}}, \emph{\bibinfo{title}{Asymptotically optimal quantum
  circuits for $d$-level systems}}, \bibinfo{journal}{Phys. Rev. Lett.}
  \textbf{\bibinfo{volume}{94}}, \bibinfo{pages}{230502}
  (\bibinfo{year}{2005}).

\bibitem[{com()}]{comment:d-wave}
\bibinfo{note}{See {http://www.dwavesys.com}}.

\bibitem[{\citenamefont{{Nigg} et~al.}(2014)\citenamefont{{Nigg}, {Mueller},
  {Martinez}, {Schindler}, {Hennrich}, {Monz}, {Martin-Delgado}, and
  {Blatt}}}]{nigg:14}
\bibinfo{author}{\bibfnamefont{D.}~\bibnamefont{{Nigg}}},
  \bibinfo{author}{\bibfnamefont{M.}~\bibnamefont{{Mueller}}},
  \bibinfo{author}{\bibfnamefont{E.~A.} \bibnamefont{{Martinez}}},
  \bibinfo{author}{\bibfnamefont{P.}~\bibnamefont{{Schindler}}},
  \bibinfo{author}{\bibfnamefont{M.}~\bibnamefont{{Hennrich}}},
  \bibinfo{author}{\bibfnamefont{T.}~\bibnamefont{{Monz}}},
  \bibinfo{author}{\bibfnamefont{M.~A.} \bibnamefont{{Martin-Delgado}}},
  \bibnamefont{and} \bibinfo{author}{\bibfnamefont{R.}~\bibnamefont{{Blatt}}},
  \emph{\bibinfo{title}{{{Experimental Quantum Computations on a Topologically
  Encoded Qubit}}}} (\bibinfo{year}{2014}),
  \bibinfo{note}{(arxiv:quant-phys/1403.5426)}.

\bibitem[{\citenamefont{Mair et~al.}(2001)\citenamefont{Mair, Vaziri, Weihs,
  and Zeilinger}}]{mair:01}
\bibinfo{author}{\bibfnamefont{A.}~\bibnamefont{Mair}},
  \bibinfo{author}{\bibfnamefont{A.}~\bibnamefont{Vaziri}},
  \bibinfo{author}{\bibfnamefont{G.}~\bibnamefont{Weihs}}, \bibnamefont{and}
  \bibinfo{author}{\bibfnamefont{A.}~\bibnamefont{Zeilinger}},
  \emph{\bibinfo{title}{{{Entanglement of the orbital angular momentum states
  of photons}}}}, \bibinfo{journal}{Nature} \textbf{\bibinfo{volume}{412}},
  \bibinfo{pages}{313} (\bibinfo{year}{2001}).

\bibitem[{\citenamefont{Piani et~al.}(2011)\citenamefont{Piani, Pitkanen,
  Kaltenbaek, and L\"utkenhaus}}]{piani:11}
\bibinfo{author}{\bibfnamefont{M.}~\bibnamefont{Piani}},
  \bibinfo{author}{\bibfnamefont{D.}~\bibnamefont{Pitkanen}},
  \bibinfo{author}{\bibfnamefont{R.}~\bibnamefont{Kaltenbaek}},
  \bibnamefont{and}
  \bibinfo{author}{\bibfnamefont{N.}~\bibnamefont{L\"utkenhaus}},
  \emph{\bibinfo{title}{Linear-optics realization of channels for single-photon
  multimode qudits}}, \bibinfo{journal}{Phys. Rev. A}
  \textbf{\bibinfo{volume}{84}}, \bibinfo{pages}{032304}
  (\bibinfo{year}{2011}).

\bibitem[{\citenamefont{Neeley et~al.}(2009)\citenamefont{Neeley, Ansmann,
  Bialczak, Hofheinz, Lucero, O'Connell, Sank, Wang, Wenner, Cleland
  et~al.}}]{neeley:09}
\bibinfo{author}{\bibfnamefont{M.}~\bibnamefont{Neeley}},
  \bibinfo{author}{\bibfnamefont{M.}~\bibnamefont{Ansmann}},
  \bibinfo{author}{\bibfnamefont{R.~C.} \bibnamefont{Bialczak}},
  \bibinfo{author}{\bibfnamefont{M.}~\bibnamefont{Hofheinz}},
  \bibinfo{author}{\bibfnamefont{E.}~\bibnamefont{Lucero}},
  \bibinfo{author}{\bibfnamefont{A.~D.} \bibnamefont{O'Connell}},
  \bibinfo{author}{\bibfnamefont{D.}~\bibnamefont{Sank}},
  \bibinfo{author}{\bibfnamefont{H.}~\bibnamefont{Wang}},
  \bibinfo{author}{\bibfnamefont{J.}~\bibnamefont{Wenner}},
  \bibinfo{author}{\bibfnamefont{A.~N.} \bibnamefont{Cleland}},
  \bibnamefont{et~al.}, \emph{\bibinfo{title}{{Emulation of a Quantum Spin with
  a Superconducting Phase Qudit}}}, \bibinfo{journal}{Science}
  \textbf{\bibinfo{volume}{325}}, \bibinfo{pages}{722} (\bibinfo{year}{2009}).

\bibitem[{\citenamefont{Klimov et~al.}(2003)\citenamefont{Klimov, Guzm\'an,
  Retamal, and Saavedra}}]{klimov:03}
\bibinfo{author}{\bibfnamefont{A.~B.} \bibnamefont{Klimov}},
  \bibinfo{author}{\bibfnamefont{R.}~\bibnamefont{Guzm\'an}},
  \bibinfo{author}{\bibfnamefont{J.~C.} \bibnamefont{Retamal}},
  \bibnamefont{and} \bibinfo{author}{\bibfnamefont{C.}~\bibnamefont{Saavedra}},
  \emph{\bibinfo{title}{{{Qutrit quantum computer with trapped ions}}}},
  \bibinfo{journal}{Phys. Rev. A} \textbf{\bibinfo{volume}{67}},
  \bibinfo{pages}{062313} (\bibinfo{year}{2003}).

\bibitem[{\citenamefont{Mischuck et~al.}(2012)\citenamefont{Mischuck, Merkel,
  and Deutsch}}]{mischuck:12}
\bibinfo{author}{\bibfnamefont{B.~E.} \bibnamefont{Mischuck}},
  \bibinfo{author}{\bibfnamefont{S.~T.} \bibnamefont{Merkel}},
  \bibnamefont{and} \bibinfo{author}{\bibfnamefont{I.~H.}
  \bibnamefont{Deutsch}}, \emph{\bibinfo{title}{Control of inhomogeneous atomic
  ensembles of hyperfine qudits}}, \bibinfo{journal}{Phys. Rev. A}
  \textbf{\bibinfo{volume}{85}}, \bibinfo{pages}{022302}
  (\bibinfo{year}{2012}).

\bibitem[{\citenamefont{Kitaev}(2003)}]{kitaev:03}
\bibinfo{author}{\bibfnamefont{A.~Y.} \bibnamefont{Kitaev}},
  \emph{\bibinfo{title}{{Fault-tolerant quantum computation by anyons}}},
  \bibinfo{journal}{Ann. Phys.} \textbf{\bibinfo{volume}{303}},
  \bibinfo{pages}{2} (\bibinfo{year}{2003}).

\bibitem[{\citenamefont{Bombin and Martin-Delgado}(2006)}]{bombin:06}
\bibinfo{author}{\bibfnamefont{H.}~\bibnamefont{Bombin}} \bibnamefont{and}
  \bibinfo{author}{\bibfnamefont{M.~A.} \bibnamefont{Martin-Delgado}},
  \emph{\bibinfo{title}{{{Topological Quantum Distillation}}}},
  \bibinfo{journal}{Phys. Rev. Lett.} \textbf{\bibinfo{volume}{97}},
  \bibinfo{pages}{180501} (\bibinfo{year}{2006}).

\bibitem[{\citenamefont{Bullock and Brennen}(2007)}]{bullock:07}
\bibinfo{author}{\bibfnamefont{S.~S.} \bibnamefont{Bullock}} \bibnamefont{and}
  \bibinfo{author}{\bibfnamefont{G.~K.} \bibnamefont{Brennen}},
  \emph{\bibinfo{title}{{{Qudit surface codes and gauge theory with finite
  cyclic groups}}}}, \bibinfo{journal}{J. Phys. A}
  \textbf{\bibinfo{volume}{40}}, \bibinfo{pages}{3481} (\bibinfo{year}{2007}).

\bibitem[{\citenamefont{Dennis et~al.}(2002)\citenamefont{Dennis, Kitaev,
  Landahl, and Preskill}}]{dennis:02}
\bibinfo{author}{\bibfnamefont{E.}~\bibnamefont{Dennis}},
  \bibinfo{author}{\bibfnamefont{A.}~\bibnamefont{Kitaev}},
  \bibinfo{author}{\bibfnamefont{A.}~\bibnamefont{Landahl}}, \bibnamefont{and}
  \bibinfo{author}{\bibfnamefont{J.}~\bibnamefont{Preskill}},
  \emph{\bibinfo{title}{{Topological quantum memory}}}, \bibinfo{journal}{J.
  Math. Phys.} \textbf{\bibinfo{volume}{43}}, \bibinfo{pages}{4452}
  (\bibinfo{year}{2002}).

\bibitem[{\citenamefont{Katzgraber et~al.}(2010)\citenamefont{Katzgraber,
  Bombin, Andrist, and Martin-Delgado}}]{katzgraber:10}
\bibinfo{author}{\bibfnamefont{H.~G.} \bibnamefont{Katzgraber}},
  \bibinfo{author}{\bibfnamefont{H.}~\bibnamefont{Bombin}},
  \bibinfo{author}{\bibfnamefont{R.~S.} \bibnamefont{Andrist}},
  \bibnamefont{and} \bibinfo{author}{\bibfnamefont{M.~A.}
  \bibnamefont{Martin-Delgado}}, \emph{\bibinfo{title}{{{Topological color
  codes on Union Jack lattices: a stable implementation of the whole Clifford
  group}}}}, \bibinfo{journal}{Phys. Rev. A} \textbf{\bibinfo{volume}{81}},
  \bibinfo{pages}{012319} (\bibinfo{year}{2010}).

\bibitem[{\citenamefont{Andrist et~al.}(2011)\citenamefont{Andrist, Katzgraber,
  Bombin, and Martin-Delgado}}]{andrist:10}
\bibinfo{author}{\bibfnamefont{R.~S.} \bibnamefont{Andrist}},
  \bibinfo{author}{\bibfnamefont{H.~G.} \bibnamefont{Katzgraber}},
  \bibinfo{author}{\bibfnamefont{H.}~\bibnamefont{Bombin}}, \bibnamefont{and}
  \bibinfo{author}{\bibfnamefont{M.~A.} \bibnamefont{Martin-Delgado}},
  \emph{\bibinfo{title}{{{Tricolored Lattice Gauge Theory with Randomness:
  Fault-Tolerance in Topological Color Codes}}}}, \bibinfo{journal}{New J.
  Phys.} \textbf{\bibinfo{volume}{13}}, \bibinfo{pages}{083006}
  (\bibinfo{year}{2011}).

\bibitem[{\citenamefont{Duclos-Cianci and Poulin}(2013)}]{duclos:13}
\bibinfo{author}{\bibfnamefont{G.}~\bibnamefont{Duclos-Cianci}}
  \bibnamefont{and} \bibinfo{author}{\bibfnamefont{D.}~\bibnamefont{Poulin}},
  \emph{\bibinfo{title}{{{Kitaev's $\mathbb{Z}_d$-code threshold estimates}}}},
  \bibinfo{journal}{Phys. Rev. A.} \textbf{\bibinfo{volume}{87}},
  \bibinfo{pages}{062338} (\bibinfo{year}{2013}).

\bibitem[{\citenamefont{Anwar et~al.}(2013)\citenamefont{Anwar, Brown,
  Campbell, and Browne}}]{anwar:13}
\bibinfo{author}{\bibfnamefont{H.}~\bibnamefont{Anwar}},
  \bibinfo{author}{\bibfnamefont{B.~J.} \bibnamefont{Brown}},
  \bibinfo{author}{\bibfnamefont{E.~T.} \bibnamefont{Campbell}},
  \bibnamefont{and} \bibinfo{author}{\bibfnamefont{D.~E.}
  \bibnamefont{Browne}}, \emph{\bibinfo{title}{{{Efficient Decoders for Qudit
  Topological Codes}}}} (\bibinfo{year}{2013}),
  \bibinfo{note}{(arXiv/quant-ph:1311.4895)}.

\bibitem[{\citenamefont{R{\"o}thlisberger
  et~al.}(2012)\citenamefont{R{\"o}thlisberger, Wootton, Heath, Pachos, and
  Loss}}]{roethlisberger:11}
\bibinfo{author}{\bibfnamefont{B.}~\bibnamefont{R{\"o}thlisberger}},
  \bibinfo{author}{\bibfnamefont{J.~R.} \bibnamefont{Wootton}},
  \bibinfo{author}{\bibfnamefont{R.~M.} \bibnamefont{Heath}},
  \bibinfo{author}{\bibfnamefont{J.~K.} \bibnamefont{Pachos}},
  \bibnamefont{and} \bibinfo{author}{\bibfnamefont{D.}~\bibnamefont{Loss}},
  \emph{\bibinfo{title}{Incoherent dynamics in the toric code subject to
  disorder}}, \bibinfo{journal}{Phys. Rev. A} \textbf{\bibinfo{volume}{85}},
  \bibinfo{pages}{022313} (\bibinfo{year}{2012}).

\bibitem[{\citenamefont{{Bombin} et~al.}(2012)\citenamefont{{Bombin},
  {Andrist}, {Ohzeki}, {Katzgraber}, and {Martin-Delgado}}}]{bombin:12}
\bibinfo{author}{\bibfnamefont{H.}~\bibnamefont{{Bombin}}},
  \bibinfo{author}{\bibfnamefont{R.~S.} \bibnamefont{{Andrist}}},
  \bibinfo{author}{\bibfnamefont{M.}~\bibnamefont{{Ohzeki}}},
  \bibinfo{author}{\bibfnamefont{H.~G.} \bibnamefont{{Katzgraber}}},
  \bibnamefont{and} \bibinfo{author}{\bibfnamefont{M.~A.}
  \bibnamefont{{Martin-Delgado}}}, \emph{\bibinfo{title}{{{Strong Resilience of
  Topological Codes to Depolarization}}}}, \bibinfo{journal}{Phys. Rev. X}
  \textbf{\bibinfo{volume}{2}}, \bibinfo{pages}{021004} (\bibinfo{year}{2012}).

\bibitem[{\citenamefont{{Nishimori}}(1981)}]{nishimori:81}
\bibinfo{author}{\bibfnamefont{H.}~\bibnamefont{{Nishimori}}},
  \emph{\bibinfo{title}{{{Internal Energy, Specific Heat and Correlation
  Function of the Bond-Random Ising Model}}}}, \bibinfo{journal}{Prog. Theor.
  Phys.} \textbf{\bibinfo{volume}{66}}, \bibinfo{pages}{1169}
  (\bibinfo{year}{1981}).

\bibitem[{\citenamefont{Hukushima and Nemoto}(1996)}]{hukushima:96}
\bibinfo{author}{\bibfnamefont{K.}~\bibnamefont{Hukushima}} \bibnamefont{and}
  \bibinfo{author}{\bibfnamefont{K.}~\bibnamefont{Nemoto}},
  \emph{\bibinfo{title}{Exchange {M}onte {C}arlo method and application to spin
  glass simulations}}, \bibinfo{journal}{J. Phys. Soc. Jpn.}
  \textbf{\bibinfo{volume}{65}}, \bibinfo{pages}{1604} (\bibinfo{year}{1996}).

\bibitem[{\citenamefont{Newman and Barkema}(1999)}]{newman:99}
\bibinfo{author}{\bibfnamefont{M.~E.~J.} \bibnamefont{Newman}}
  \bibnamefont{and} \bibinfo{author}{\bibfnamefont{G.~T.}
  \bibnamefont{Barkema}}, \emph{\bibinfo{title}{{M}onte {C}arlo Methods in
  Statistical Physics}} (\bibinfo{publisher}{Oxford University Press Inc.},
  \bibinfo{address}{New York, USA}, \bibinfo{year}{1999}).

\bibitem[{\citenamefont{Katzgraber et~al.}(2006)\citenamefont{Katzgraber,
  Trebst, Huse, and Troyer}}]{katzgraber:06a}
\bibinfo{author}{\bibfnamefont{H.~G.} \bibnamefont{Katzgraber}},
  \bibinfo{author}{\bibfnamefont{S.}~\bibnamefont{Trebst}},
  \bibinfo{author}{\bibfnamefont{D.~A.} \bibnamefont{Huse}}, \bibnamefont{and}
  \bibinfo{author}{\bibfnamefont{M.}~\bibnamefont{Troyer}},
  \emph{\bibinfo{title}{{{Feedback-optimized parallel tempering Monte
  Carlo}}}}, \bibinfo{journal}{J. Stat. Mech.}
  \textbf{\bibinfo{volume}{\normalfont{P03018}}} (\bibinfo{year}{2006}).

\bibitem[{\citenamefont{Palassini and Caracciolo}(1999)}]{palassini:99b}
\bibinfo{author}{\bibfnamefont{M.}~\bibnamefont{Palassini}} \bibnamefont{and}
  \bibinfo{author}{\bibfnamefont{S.}~\bibnamefont{Caracciolo}},
  \emph{\bibinfo{title}{{U}niversal {F}inite-{S}ize {S}caling {F}unctions in
  the 3{D} {I}sing {S}pin {G}lass}}, \bibinfo{journal}{Phys. Rev. Lett.}
  \textbf{\bibinfo{volume}{82}}, \bibinfo{pages}{5128} (\bibinfo{year}{1999}).

\bibitem[{\citenamefont{Katzgraber et~al.}(2009)\citenamefont{Katzgraber,
  Bombin, and Martin-Delgado}}]{katzgraber:09c}
\bibinfo{author}{\bibfnamefont{H.~G.} \bibnamefont{Katzgraber}},
  \bibinfo{author}{\bibfnamefont{H.}~\bibnamefont{Bombin}}, \bibnamefont{and}
  \bibinfo{author}{\bibfnamefont{M.~A.} \bibnamefont{Martin-Delgado}},
  \emph{\bibinfo{title}{{Error Threshold for Color Codes and Random 3-Body
  Ising Models}}}, \bibinfo{journal}{Phys. Rev. Lett.}
  \textbf{\bibinfo{volume}{103}}, \bibinfo{pages}{090501}
  (\bibinfo{year}{2009}).

\bibitem[{\citenamefont{Thomas and Katzgraber}(2011)}]{thomas:11d}
\bibinfo{author}{\bibfnamefont{C.~K.} \bibnamefont{Thomas}} \bibnamefont{and}
  \bibinfo{author}{\bibfnamefont{H.~G.} \bibnamefont{Katzgraber}},
  \emph{\bibinfo{title}{Simplest model to study reentrance in physical
  systems}}, \bibinfo{journal}{Phys. Rev. E} \textbf{\bibinfo{volume}{84}},
  \bibinfo{pages}{040101(R)} (\bibinfo{year}{2011}).

\end{thebibliography}

\end{document}